\documentclass{aa}
\usepackage{graphicx}
\usepackage[varg]{txfonts}
\usepackage{cases}
\usepackage{natbib}

\begin{document}
   \title{Transverse coronal loop oscillations excited by homologous circular-ribbon flares}

   \author{Q. M. Zhang\inst{1,2,3}, J. Dai\inst{1}, Z. Xu\inst{4}, D. Li\inst{1}, L. Lu\inst{1}, K. V. Tam\inst{2}, and A. A. Xu\inst{2}}

   \institute{Key Laboratory of Dark Matter and Space Science, Purple Mountain Observatory, CAS, Nanjing 210033, PR China \\
              \email{zhangqm@pmo.ac.cn}
              \and
              State Key Laboratory of Lunar and Planetary Sciences, Macau University of Science and Technology, Macau, PR China \\
              \and
              School of Astronomy and Space Science, Nanjing University, Nanjing 210023, PR China \\
              \and
              Yunnan Observatories, Chinese Academy of Sciences, Kunming 650011, PR China \\
              }

   \date{Received; accepted}
    \titlerunning{Transverse coronal loop oscillations}
    \authorrunning{Zhang et al.}

 \abstract
   {}
   {We report our multiwavelength observations of two homologous circular-ribbon flares (CRFs) in active region 11991 on 2014 March 5, 
   focusing on the transverse oscillations of an extreme-ultraviolet (EUV) loop excited by the flares.}
   {The flares were observed in ultraviolet (UV) and EUV wavelengths by the Atmospheric Imaging Assembly (AIA) 
   on board the Solar Dynamics Observatory (SDO) spacecraft. They were also observed in H$\alpha$ line center by the one-meter New Vacuum Solar Telescope (NVST).
   Soft X-ray (SXR) fluxes of the flares in 0.5$-$4 and 1$-$8 {\AA} were recorded by the GOES spacecraft.}
   {The transverse oscillations are of fast standing kink-mode.
   The first-stage oscillation triggered by the C2.8 flare is decayless with lower amplitudes (310$-$510 km). 
   The periods (115$-$118 s) in different wavelengths are nearly the same, indicating coherent oscillations.
   The magnetic field of the loop is estimated to be 65$-$78 G. 
   The second-stage oscillation triggered by the M1.0 flare is decaying with larger amplitudes (1250$-$1280 km). 
   The periods decreases from 117 s in 211 {\AA} to 70 s in 171 {\AA}, implying a decrease of loop length or an implosion after a gradual expansion.
   The damping time, being 147$-$315 s, increases with the period, so that the values of $\tau/P$ are close to each other in different wavelengths.
   The thickness of the inhomogeneous layer is estimated to be $\sim$0\farcs45 under the assumption of resonant absorption.}
   {This is the first observation of the excitation of two kink-mode loop oscillations by two sympathetic flares.
   The results are important for understanding of the excitation of kink oscillations of coronal loops and hence the energy balance in the solar corona.
   Our findings also validate the prevalence of significantly amplified amplitudes of oscillations by successive drivers.}


 \maketitle

\section{Introduction} \label{s-intro}
Waves and oscillations are ubiquitous in the solar atmosphere \citep[see][and references therein]{naka05,wang11,naka16}. They are widely detected in sunspot \citep{tian14}, spicules \citep{dep12}, 
coronal jets \citep{cir07}, prominences \citep{oka07,zqm12b,shen14,zqm17,luna18}, polar plumes \citep{def98}, coronal loops \citep{wang04,go12,kim14,li20}, and hot post flare loops \citep{tian16}.
The magnetohydrodynamic (MHD) waves can be classified into fast mode waves \citep{chen11,zim15}, slow mode waves \citep{ofm02,wang03}, and Alfv\'{e}n waves \citep{ban98,erd07}.
Coronal loop oscillations excited by solar flares are first discovered by the TRACE mission \citep{asch99,naka99}. The initial amplitudes range from a few to 30 megameter (Mm). 
The periods (2$-$20 minutes) are found to be proportional to the loop lengths \citep{god16b}. In most cases, the transverse oscillations are of standing kink mode \citep{naka01,asch02,ver04}, 
which provides a useful tool to estimate the magnetic field and Alfv\'{e}n speed of the coronal loops \citep{wang02,van08,ver09,wv12,ver13,yuan16,li17}. 
Sometimes higher harmonics could be detected besides the fundamental mode \citep{dem07,van07,wht12,guo15}. 
The period ratio is applied to estimate the density scale height \citep{van09,duck18}.

The amplitude of transverse loop oscillations usually attenuates with time, with the exponential damping time ($\tau$) being 2$-$40 minutes \citep{god16a}. A linear fit between the damping time and 
period ($P$) results in $\tau=(1.53\pm0.03)P$ \citep{god16b}. Resonant absorption plays an important role in the rapid damping of kink oscillations, which acts in a finite inhomogenous layer 
of a flux tube and leads to a transfer of energy from kink to Alfv\'{e}n mode oscillation \citep{goo02,rud02}. The kink mode damping rate provides a powerful diagnostic tool to estimate the coronal loop
density profile \citep{pas16}. Recently, 3D numerical simulations and forward modeling of standing transverse MHD waves in coronal loops reveal that the observed signatures are dominated by the 
combination of the Kelvin-Helmholtz instability (KHI), resonant absorption, and phase mixing \citep{ant17}. 
Transverse oscillations with growing amplitudes have occasionally been identified \citep{wang12}.
Some of the low-amplitude ($\la 0.5$ Mm) oscillations hardly attenuate with time, which are termed decay-less oscillations \citep{anf13,anf15,li18,afa20}.
\citet{nis13} studied an eruptive flare in active region (AR) 11494 on 2012 May 30. Before and well after the occurrence of flare, the coronal loops in the same AR experienced low-amplitude 
decayless oscillations, while large-amplitude oscillations triggered by the flare decayed with time. So far, transverse oscillations triggered by successive flares have rarely been reported.

In this paper, we revisit the homologous C2.8 and M1.0 circular-ribbon flares (CRFs) on 2014 March 5 \citep{xu17}, focusing on the transverse coronal loop oscillations excited by the flares.
In Sect.~\ref{s-data}, we describe the data analysis. Results are presented in Sect.~\ref{s-res}. 
We compare our findings with previous results in Sect.~\ref{s-disc} and give a summary in Sect.~\ref{s-sum}.

\section{Observations and data analysis} \label{s-data}
The confined flares in NOAA AR 11991 (S24W25) were observed by the Atmospheric Imaging Assembly \citep[AIA;][]{lem12} on board the Solar Dynamics Observatory (SDO).
AIA takes full-disk images in two ultraviolet (UV; 1600 and 1700 {\AA}) and seven extreme-ultraviolet (EUV; 94, 131, 171, 193, 211, 304, and 335 {\AA}) wavelengths.
The level\_1 data were calibrated using the standard solar software (SSW) program \texttt{aia\_prep.pro}.
The flares were also observed in H$\alpha$ line center by the one-meter New Vacuum Solar Telescope \citep[NVST;][]{liu14} located at the Fuxian Solar Observatory.
The raw data were reconstructed by a speckle masking method following the flat-field and dark-field processing \citep{xu14}.
The level\_1.5 H$\alpha$ images coaligned with the 1600 {\AA} images were used for analysis.
Soft X-ray (SXR) light curves of the flares in 0.5$-$4 and 1$-$8 {\AA} were recorded by the GOES spacecraft. Similar to the C3.1 CRF on 2015 October 16 \citep{zqm16}, 
the two flares were associated with two type III radio bursts, which were evident in the radio dynamic spectra by WIND/WAVES \citep{bou95}.
The observational parameters during 01:30$-$02:20 UT are listed in Table~\ref{tab-1}.

\begin{table}
\centering
\caption{Description of the observational parameters.}
\label{tab-1}
\begin{tabular}{cccc}
\hline\hline
Instrument & $\lambda$   &  Cad. & Pix. Size \\ 
                  & ({\AA})         &   (s)           & (\arcsec) \\
\hline
SDO/AIA & 171$-$211 &  12 & 0.6 \\
SDO/AIA & 1600        &  24 & 0.6 \\
NVST      & 6562.8     &  12 & 0.165 \\
GOES     & 0.5$-$4.0 &  2.05 & ... \\
GOES     & 1$-$8    &  2.05 & ... \\
WIND/WAVES      & 0.02$-$13.825 MHz &  60 & ... \\
\hline
\end{tabular}
\end{table}

\section{Results} \label{s-res}
\begin{figure}
\includegraphics[width=9cm,clip=]{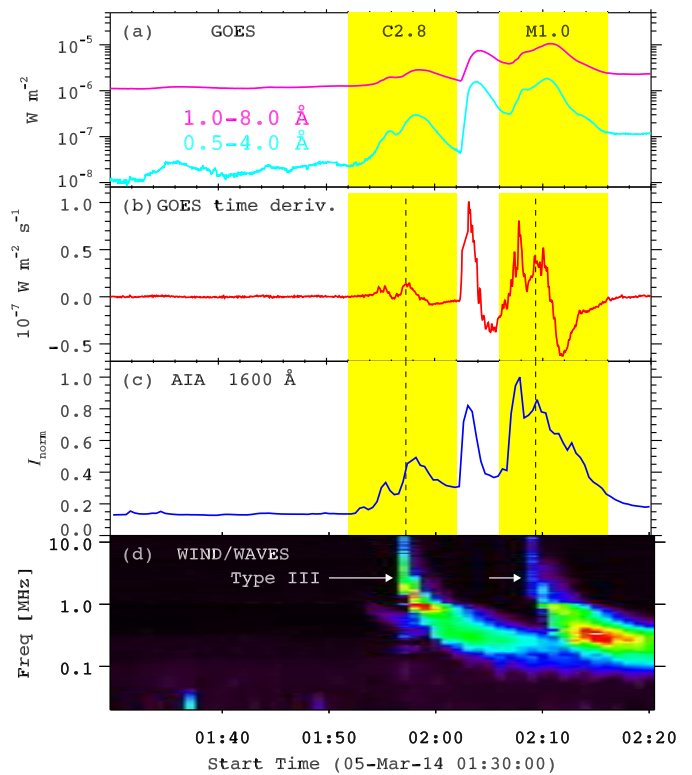}
\centering
\caption{(a) SXR light curves of the homologous flares in 0.5$-$4 {\AA} (cyan line) and 1$-$8 {\AA} (magenta line).
(b) Time derivative of the 1$-$8 {\AA} flux.
(c) Light curve of the flares in AIA 1600 {\AA}. The yellow regions stand for the times of CRFs.
(d) Radio dynamic spectra recorded by WIND/WAVES. The arrows point to the type III radio bursts.}
\label{fig1}
\end{figure}

\begin{figure}
\includegraphics[width=9cm,clip=]{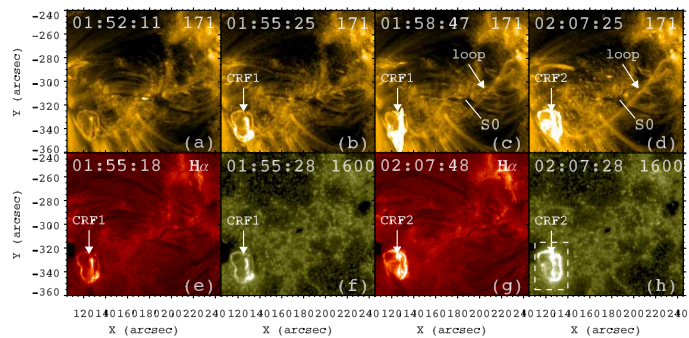}
\centering
\caption{Snapshots of the AIA 171 {\AA}, 1600 {\AA}, and H$\alpha$ images.
The arrows point to CRF1, CRF2, and the EUV loop that experiences kink oscillations.
The slice (S0) is used to investigate the loop oscillations.
The dashed box in panel (h) signifies the flare region.
The whole evolution is shown in a movie (\textit{anim171.mov}) that is available online.}
\label{fig2}
\end{figure}

In Fig.~\ref{fig1}, the top panel shows SXR light curves of the flares. 
The SXR emissions of the C2.8 flare (CRF1) started to rise at $\sim$01:52 UT and reached peak values at $\sim$01:58 UT, which were followed by a gradual decay phase until $\sim$02:02 UT. 
The SXR emissions of the M1.0 flare (CRF2) started to rise at $\sim$02:06 UT and reached peak values at $\sim$02:10 UT, which were followed by a decay phase until $\sim$02:16 UT. 
Hence, the lifetimes of CRF1 and CRF2 are only $\sim$10 minutes. 
Figure~\ref{fig1}(b) shows the time derivative of the 1$-$8 {\AA} flux, which serves as a hard X-ray (HXR) proxy according to the Neupert effect.  
Light curve of the flares in AIA 1600 {\AA}, defined as the integral intensities of the flare region in Fig.~\ref{fig2}(h), is plotted in Fig.~\ref{fig1}(c).
It is clearly seen that the major peaks in SXR derivative and 1600 {\AA} have almost one-to-one correspondence during the two flares.
Radio dynamic spectra recorded by WIND/WAVES are drawn in Fig.~\ref{fig1}(d), featuring two type III radio bursts starting at $\sim$01:57 UT and $\sim$02:09 UT, respectively. 
The occurrence of type III radio burst is an indication of outward propagating nonthermal electrons accelerated by flares along open magnetic field lines (see also Fig. 8 in \citet{xu17}).
The coincidence between the starting times of radio bursts and peaks in UV/HXR suggests bidirectional nonthermal electrons propagating upward along open field and downward 
along reconnected field into the chromosphere.

In Fig.~\ref{fig2}, the top panels show snapshots of the AIA 171 {\AA} images (see also the online movie \textit{anim171.mov}). 
Panel (a) shows the image at the very beginning of CRF1. The brightness of flare ribbons, including a short inner ribbon (IR) and an outer circular ribbon (CR), reached their peak values 
simultaneously in EUV (see panels (b-c)), UV (see panel (f)), and H$\alpha$ (see panel (e)) wavelengths. The size ($\sim$26$\arcsec$) of the CR is comparable to that of jet-related coronal 
bright points \citep{zqm12a}. For CRF2, the brightness of flare ribbons reached their peak values at $\sim$02:07 UT (see panels (d), (g), and (h)). It is noted that both flares were associated
with cool surges propagating in the southwest direction (see Fig. 3(h) and Fig. 4(k) in \citet{xu17}). 

\begin{figure}
\includegraphics[width=9cm,clip=]{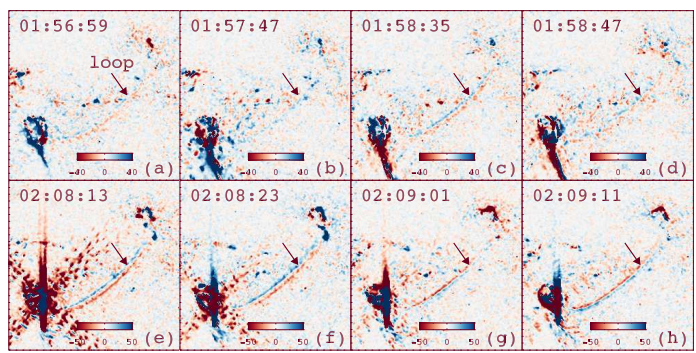}
\centering
\caption{Running-difference images in 211 {\AA}, where blue (red) color represents intensity enhancement (weakening).
Evolution of the kink oscillations of the EUV loop is shown in a movie (\textit{anim211.mov}) that is available online.}
\label{fig3}
\end{figure}

As mentioned in \citet{xu17}, a long EUV loop connecting the flares with remote brightenings (RB), showed up during the flares, which are pointed by arrows in Fig.~\ref{fig2}(c-d). 
The loop experienced transverse oscillations excited by the homologous flares. To better illustrate the displacements of the loop, we apply the running difference technique to the original EUV images.
Running-difference images in 211 {\AA} during CRF1 and CRF2 are displayed in the top and bottom panels of Fig.~\ref{fig3} (see also the online movie \textit{anim211.mov}).
The blue (red) color represents intensity enhancement (weakening), respectively. It is obviously revealed that the loop oscillated back and forth in a coherent way and the 
displacements of different segments of the loop are in-phase, which is indicative of fast standing kink-mode oscillations.

\begin{figure}
\includegraphics[width=9cm,clip=]{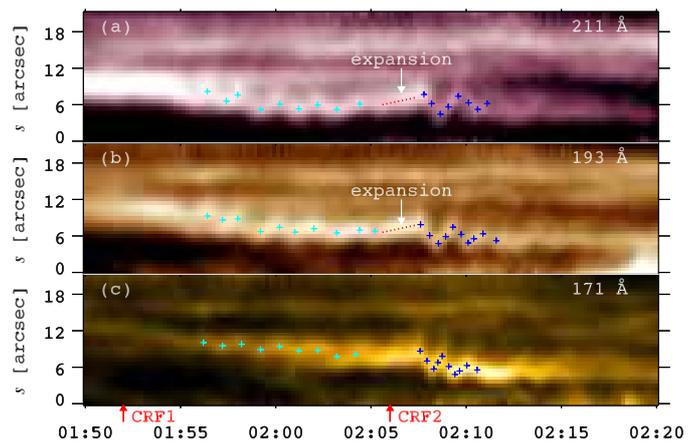}
\centering
\caption{Time-distance diagrams of S0 in different wavelengths, showing the transverse oscillations of the EUV loop.
The cyan (blue) plus symbols denote the central positions of the loop excited by CRF1 (CRF2).
The red dotted lines indicate slow expansion of the loop before CRF2.
The red arrows on the $x$-axis signify the starting times of CRF1 and CRF2 in SXR.
$s=0$ and $s=21\arcsec$ on the $y$-axis denote the northeast and southwest endpoints of S0, respectively.}
\label{fig4}
\end{figure}

The oscillations could excellently be recognized in AIA 171, 193, and 211 {\AA}. 
To investigate the characteristics of kink oscillations, we select an artificial slice (S0) with a length of 21$\arcsec$ across the loop and close to the loop apex (see Fig.~\ref{fig2}(d)).
Time-distance diagrams of S0 in different wavelengths are displayed in Fig.~\ref{fig4}. The cyan and blue symbols denote the central positions of the loop.
The first-stage small-amplitude oscillation excited by CRF1 and second-stage large-amplitude oscillation excited by CRF2 are distinctly demonstrated. 

\begin{table*}
\centering
\caption{Parameters of coronal loop oscillations observed by AIA in different wavelengths. 
$b_1$ and $b_2$ stand for the drift speeds of the loop.}
\label{tab-2}
\begin{tabular}{c|ccc|ccccc}
\hline\hline
$\lambda$ & $b_1$ & $A_1$ & $P_1$ & $b_2$ & $A_2$ & $P_2$ & $\tau_2$ & $\tau_{2}/P_{2}$ \\
({\AA})         & (km s$^{-1}$) & (km) & (s) & (km s$^{-1}$) & (km) & (s) & (s) & \\
\hline
171 & 3.1 & 329$\pm$82 & 118$\pm$5 & 6.6 & 1254$\pm$122 &  70$\pm$3 & 147$\pm$18 & 2.1 \\
193 & 2.6 & 312$\pm$77 & 115$\pm$4 & 1.4 & 1280$\pm$129 & 100$\pm$4 & 257$\pm$27 & 2.6 \\
211 & 2.5 & 510$\pm$110 & 117$\pm$5 & 1.2 & 1283$\pm$130 & 117$\pm$5 & 315$\pm$30 & 2.7 \\
\hline
ave. & 2.7 & 383 & 116.7 & 3.1 & 1272 & 95.7 & 240 & 2.5 \\
\hline
\end{tabular}
\end{table*}

\begin{figure}
\includegraphics[width=9cm,clip=]{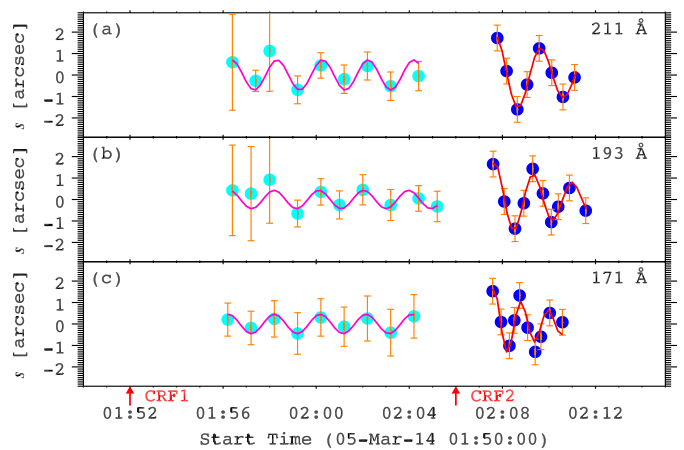}
\centering
\caption{Detrended central positions of the EUV loop in different wavelengths. 
Oscillations excited by CRF1 (cyan circles) and by CRF2 (blue circles) are fitted with a decayless (magenta lines) and an exponentially decaying (red lines) sine functions, respectively.}
\label{fig5}
\end{figure}

It is noticed that the loop drifted nonlinearly in the northeast direction during the oscillations, which is probably related to the counterclockwise motion of the bright CR. 
The apparent drift speeds, ranging from $\sim$1 to $\sim$6 km s$^{-1}$ with an average value of $\sim$3 km s$^{-1}$, are listed in the second and fifth columns of Table~\ref{tab-2}.
The detrended central positions of the EUV loop in different wavelengths are plotted in Fig.~\ref{fig5}.
Oscillation excited by CRF1 (cyan circles) started at $\sim$01:56 UT and lasted for $\sim$10 minutes. 
It is fitted with a decayless sine function (magenta lines) using the standard SSW program \texttt{mpfit.pro}:
\begin{equation} \label{eqn-1}
  y_{1}=A_{1}\sin(\frac{2\pi}{P_1}t+\phi_1),
\end{equation}
where $A_{1}$, $\phi_1$, and $P_1$ stand for the initial amplitude, initial phase, and period, respectively.
The derived values of $A_{1}$ and $P_1$ are listed in the third and fourth columns of Table~\ref{tab-2}.
It is seen that the amplitude increases from $\sim$310 km in 193 {\AA} to $\sim$510 km in 211 {\AA}, with an average value of $\sim$383 km. 
The periods (115$-$118 s) are almost the same, suggesting that the loop oscillates in phase with low amplitudes in different wavelengths.

Oscillation excited by CRF2 (blue circles) started at $\sim$02:07:30 UT and lasted for $\sim$4 minutes. It is fitted with an exponentially decaying sine function (red lines):
\begin{equation} \label{eqn-2}
  y_{2}=A_{2}\sin(\frac{2\pi}{P_2}t+\phi_2)e^{-t/\tau_{2}},
\end{equation}
where $A_{2}$, $\phi_2$, $P_2$, and $\tau_{2}$ stand for the initial amplitude, initial phase, period, and damping time, respectively.
The derived values of $A_{2}$, $P_2$, $\tau_{2}$, and $\tau_{2}/P_{2}$ are listed in the last four columns of Table~\ref{tab-2}.
It is revealed that the initial amplitudes of transverse oscillation excited by CRF2 are significantly larger than the amplitudes excited by CRF1.
In other words, the low-amplitude loop oscillation is remarkably amplified or enhanced by the second flare.
The period remains unchanged in 211 {\AA}, while decreases by 15 s in 193 {\AA} and decreases considerably by 48 s in 171 {\AA}, 
meaning that the large-amplitude oscillation is no longer coherent in different wavelengths.
The damping time ranges from 147 to 315 s, whereas the values of $\tau_{2}/P_{2}$ are close to each other.

In Fig.~\ref{fig4}(a-b), a slow expansion (red dotted lines) is found before the large-amplitude oscillation, which is consistent with the gradual expansion phase 
before the main collapse and oscillation phase \citep{sim13}. Assuming that the phase speeds of the decaying and decayless kink-mode oscillations are equivalent, 
the decrease of periods in 171 and 193 {\AA} may suggest a decrease of loop length, or an implosion.

\section{Discussion} \label{s-disc}
Since the first detection of transverse coronal loop oscillations \citep{asch99}, there are abundant observations and numerical modelings. 
Coronal seismology becomes a powerful approach to diagnose the magnetic field strength of the oscillating loops, which is hard to measure directly \citep{and09}.
For the standing kink-mode oscillations of the EUV loop, which corresponds to the outer spine connecting the null point with RB \citep{xu17},
the period depends on the loop length ($L$) and phase speed ($C_k$) \citep{naka99}:
\begin{equation} \label{eqn-3}
  P=2L/C_k, C_k=\sqrt{2/(1+\rho_o/\rho_i)}C_{A},
\end{equation}
where $C_{A}$ is the Alfv\'{e}n speed of the loop, and $\rho_i$ and $\rho_o$ stand for the internal and external plasma densities.
The value of $L$ is estimated to be $\sim$130 Mm, assuming a semicircular shape. $C_k$ is estimated to be $\sim$2200 km s$^{-1}$ by adopting $P=117$ s.
Hence, $C_{A}\approx1555$ km s$^{-1}$, assuming that the density ratio $r=\rho_o/\rho_i\approx0.1$ \citep{naka99}.
The magnetic field of a transversely oscillating loop is expressed as \citep{nis13}:
\begin{equation} \label{eqn-4}
  B=C_{A}\sqrt{4\pi\mu_{C}\rho_i},
\end{equation}
where $\mu_{C}=1.27$ is the average molecular weight in the corona.
Taking the number density of the loop to be (7$-$10)$\times$10$^9$ cm$^{-3}$ \citep{sun13}, $B$ is in the range of 65$-$78 G.

As shown in Fig.~\ref{fig5}, the large-amplitude kink oscillation of the EUV loop damps rapidly with time. In the case of resonant absorption \citep{rud02}, $\tau/P$ is expressed as:
\begin{equation} \label{eqn-4}
  \frac{\tau}{P}=\frac{2a}{\pi l}\frac{1+r}{1-r},
\end{equation}
where $a\approx1\farcs5$ is the loop half-width and $l$ is the thickness of the inhomogeneous layer.
Based on the measured $\tau/P$ in Table~\ref{tab-2}, $l$ is estimated to be $\sim$0\farcs45. The ratio $l/a\approx0.3$ is close to the value reported by \citet{ver09}.

So far, the observations of transverse loop oscillations excited by successive flares are rare. For the first time, \citet{nis13} reported decayless low-amplitude oscillation followed by decaying 
high-amplitude oscillation, which is interpreted by a damped linear oscillator excited by a continuous low-amplitude harmonic driver and by an impulsive high-amplitude driver (e.g., a flare).
\citet{kum13} reported the kink oscillations of a coronal loop initially driven by a fast-mode EUV wave and later amplified by a slower EIT wave.
For the first time, \citet{her11} investigated two successive trains of large-amplitude transverse oscillations in an EUV prominence 
excited by two coronal waves associated with two sympathetic flares.
Enhancement of amplitudes by homologous confined flares has also been discovered in longitudinal filament oscillations \citep{zqm20}.
In our case, the EUV loop undergoes two-stage transverse oscillations: decayless low-amplitude oscillation excited by the C2.8 flare, 
and decaying high-amplitude oscillation excited by the M1.0 flare, which is 14 minutes later than the first one.
Therefore, noticeable amplification of amplitudes by successive drivers is prevalent not only in coronal loop oscillations but also in filament oscillations.

It should be emphasized that the apparent decayless oscillation excited by the first C2.8 flare might be different from the regular decayless oscillations, 
which appear without any flares \citep{nis13,afa20}. Alternatively, the first flare could cause a turn of the oscillating loop, making the plane of the oscillation polarization 
closer to the plane of the sky, and hence making the decay-less oscillations more visible. In Fig.~\ref{fig1}(a), there is a clear SXR peak between the C2.8 and M1.0 flares. 
The emissions of the peak come from the same AR 11991. However, it is not recorded as an independent flare\footnote{https://solarmonitor.org}. 
The low-amplitude decayless oscillation of the EUV loop is probably affected by the energy release as a non-resonant external force \citep{anf13}, 
so that it continues till the excitation of the large-amplitude oscillation by the second flare (see Fig.~\ref{fig4}).

\section{Summary} \label{s-sum}
In this work, we report our multiwavelength observations of two homologous CRFs observed by SDO/AIA and NVST on 2014 March 5. 
Both of them excited transverse kink-mode oscillations of an EUV loop that corresponds to the outer spine connecting the null point with RB.
The first-stage oscillation triggered by the C2.8 flare is decayless with lower amplitudes (310$-$510 km). 
The periods (115$-$118 s) in different wavelengths are nearly the same, indicating coherent oscillations.
The magnetic field of the loop is estimated to be 65$-$78 G.
The second-stage oscillation triggered by the M1.0 flare is decaying with larger amplitudes (1250$-$1280 km). 
The periods decreases from 117 s in 211 {\AA} to 70 s in 171 {\AA}, implying a decrease of loop length or an implosion after a slow expansion.
The damping time, being 147$-$315 s, increases with the period, so that the values of $\tau/P$ are close to each other in different wavelengths.
The thickness of the inhomogeneous layer is estimated to be $\sim$0\farcs45 under the assumption of resonant absorption.
The results are important for our understanding of the excitation of kink oscillations of coronal loops and the energy balance in the corona.
Our findings also validate the prevalence of significantly amplified amplitudes of oscillations by successive drivers.

\begin{acknowledgements}
The authors appreciate the referee for valuable suggestions to improve the quality of this paper.
SDO is a mission of NASA\rq{}s Living With a Star Program. AIA data are courtesy of the NASA/SDO science teams.
This work is funded by NSFC grants (No. 11773079, 11790302, 11873091, 11973092), the International Cooperation and Interchange Program (11961131002), 
the Youth Innovation Promotion Association CAS, Yunnan Province Basic Research Plan (No. 2019FA001), the Science and Technology Development Fund of Macau (275/2017/A), 
CAS Key Laboratory of Solar Activity, National Astronomical Observatories (KLSA202003, KLSA202006), 
the Strategic Priority Research Program on Space Science, CAS (XDA15052200, XDA15320301), 
and the project supported by the Specialized Research Fund for State Key Laboratories.
\end{acknowledgements}

\end{document}